\documentclass[fleqn,usenatbib]{mnras}

\usepackage{newtxtext,newtxmath}

\usepackage[T1]{fontenc}
\usepackage{ae,aecompl}

\usepackage{graphicx}	
\usepackage{amsmath}	
\usepackage{amssymb}	
\usepackage{float}

\title[galaxy-halo relation]{Modelling the tightest relation between galaxy properties and dark matter halo properties from hydrodynamical simulations of galaxy formation}

\author[Jian-hua He]{
Jian-hua He,$^{1,2}$\thanks{E-mail: hejianhua@nju.edu.cn}
\\
$^{1}$School of Astronomy and Space Science, Nanjing University, Nanjing 210093, P. R. China\\
$^{2}$Institute for Computational Cosmology, Department of Physics, Durham University, South Road, Durham DH1 3LE, UK
\\
}

\date{Accepted XXX. Received YYY; in original form ZZZ}

\pubyear{2019}

\begin{document}
\label{firstpage}
\pagerange{\pageref{firstpage}--\pageref{lastpage}}
\maketitle

\begin{abstract}
We investigate how a property of a galaxy correlates most tightly with a property of its host dark matter halo, using state-of-the-art hydrodynamical simulations of galaxy formation EAGLE, Illustris, and IllustrisTNG. Unlike most of the previous work, our analyses focus on all types of galaxies, including both central and satellite galaxies. We find that the stellar mass of a galaxy at the epoch of the peak circular velocity with an evolution correction gives the tightest such correlation to the peak circular velocity $V_{\rm peak}$ of the galaxy's underling dark matter halo. The evolution of galaxy stellar mass reduces rather than increases scatter in such a relation. We also find that one major source of scatter comes from star stripping due to the strong interactions between galaxies. Even though, we show that the size of scatter predicted by hydrodynamical simulations has a negligible impact on the clustering of dense $V_{\rm peak}$-selected subhalo from simulations, which suggests that even the simplest subhalo abundance matching (SHAM), without scatter and any additional free parameter, can provide a robust prediction of galaxy clustering that can agree impressively well with the observations from the SDSS main galaxy survey.
\end{abstract}



\section{Introduction}

In modern analysis of the large-scale structure of galaxies, galaxies are usually treated as discrete points. The over-density field of galaxies is a Poisson sample of the underlying dark matter field~\citet{Peebles} and the statistics of galaxies does not depend on properties of galaxies. However, this assumption is indeed over-simplified~\citet{CasasMiranda:2001ym}. Galaxy formation is far more complicated than just a discrete point process. Galaxies are not formed in an isolated environment but rather interact frequently with one another. Not only can such interactions change the probability distribution of galaxies but also the properties of galaxy themselves. As a result, galaxies are biased tracers of the dark matter field. The way they trace the dark matter field depends both on their properties as well as their assembly histories. For instance, it is well known that red galaxies are more clustered than the blue ones; more luminous galaxies are more clustered than the fainter ones as well~\citet{Idit}. Even for galaxies with similar properties, such as those selected in the BOSS CMASS sample~\citet{Dawson_2012}, where the galaxies have a fairly uniform mass distribution that peaks at around $\log(M/M_{\odot})\sim11.3$, galaxy clustering is still dependent on the stellar-mass assembly history~\citet{Montero-Dorta:2017hsx}.

Since in the realistic case galaxies are not good tracers of the dark matter field, the key question then becomes how to accurately quantify the relation between them. The first approach for this is to do modeling. In $\Lambda$CDM, there has been significant progresses in modeling such biases over the past decade~\citet{Scoccimarro:2004tg,TNS,delaTorre:2016rxm,Sylvain,BianchiII,Bianchi:2014kba}. The galaxy bias model now can yield reasonable accuracy in the quasi-linear regime. However, despite the progress, this approach is still unsatisfactory in several aspects. First, the accuracy and utility of these models have to be tested against mock galaxy surveys, as the galaxy bias models are essentially phenomenological and empirical. However, the mock catalogues are usually built for a particular survey or specific to a particular selection of galaxies. It is unclear how accurate these bias models can be applied to samples with a wide range of properties and complex biases. Moreover, it also remains difficult to assess the extent to which the underlining simplified assumptions and intrinsic limitations in the mock catalogues affect the calibration of these models in the first place. In addition, aside from the accuracy and utility issue, another concern is that current bias models are tested against mock galaxy catalogues based on simulations only in $\Lambda$CDM. It is unknown whether or not these models can still work in a modified gravity model. This is an important test because, unlike in $\Lambda$CDM, the relationship between galaxies and the dark matter field is much more complicated in modified gravity (MG) models. Galaxies in MG models are not even directly related to the dark matter field, as the formation, clustering, and motion of galaxies are dictated by a different potential $\phi_+=(\psi+\phi)/2$ rather than the lensing potential $\phi_-=(\psi-\phi)/2$ that is directly related to the true dark matter field. The relationship between $\phi_+=(\psi+\phi)/2$ and $\phi_-=(\psi-\phi)/2$ can be very complex~\citet{He:2015bua} in a modified gravity model.

The second approach makes use of N-body simulations, coupled to some phenomenological frameworks, such as the halo occupation distribution (HOD)~\citet{Jing:1997nb,Peacock:2000qk,Berlind:2001xk,Zheng:2004id,Zheng:2007zg,Leauthaud_2011,Idit,Guo:2015dda,Guo:2015epa} or the conditional luminosity function (CLF)~\citet{Yang:2002ww,Yang:2008eg}, to link galaxies to dark matter halos. The basic idea of this approach is that galaxies reside in dark matter halos, namely, the densest regions of the underlying dark matter field. And the probability of the distribution of galaxies is only dependent on the masses of dark matter halos. Although these assumptions are obviously over-simplified which neglects some important effects such as the assembly bias~\citet{Gao:2005ca,Gao:2006qz}, the HOD/CLF modeling turns out to be very successful in reproducing galaxy clustering even at very small scales for galaxy samples with a wide variety of different properties and complex biases~\citet{Idit,Zheng:2007zg,Guo:2015dda,Guo:2015epa}. However, despite this, when applied to testing the underlining cosmological models, these frameworks have obvious limitations: they are too flexible and, in general, lack of strong physical motivations. In a modified gravity model, for example, even naively using the same HOD parameterisations as in $\Lambda$CDM, the framework can still be tuned to reproduce desired galaxy clustering~\citet{Hernandez-Aguayo:2018oxg}, which is clearly unfeasible given that the dark matter halo properties and the processes of galaxy formation in a modified gravity model ought to be very different from those in $\Lambda$CDM.

In this work, rather than adhering to the doctrine that galaxies are tracers of the dark matter field, we regard galaxies as tracers of the dark matter halos: galaxies illuminate the dark matter halos and their properties are tracers of the properties of the dark matter halos. This idea is in line with the philosophy of subhalo abundance matching~\citet{Kravtsov:2003sg,Vale:2004yt,Conroy:2005aq,moster,qiguo,Behroozi_2013,Reddick_2014}. In this paper, we will further explore how a property of galaxy physically correlates to a property of its host dark matter halo. In order to achieve this, we use hydro-dynamical simulations of galaxy formation. In contrast to the sim-analytical galaxy formation models that are based on the DMO ones~\citet{Cole:2000ex,guoqi}, hydrodynamical simulations can trace the complicated interactions between baryons and dark matter in a self-consistent way based on the first principles of gravity and hydrodynamics. This is of paramount importance as the motion and clustering of galaxies, along with their host dark matter halos, are primarily dictated by gravity and hydrodynamics. In order to further strength our results and prevent potential biases due to the choice of a particular simulation, in this work we adopt three different simulations EAGLE~\citet{Schaye:2014tpa,EAGLE2,McAlpine:2015tma}, Illustris~\citet{Nelson:2015dga,Vogelsberger:2014dza} and IllustrisTNG ~\cite{Nelson:2018uso,Pillepich:2017fcc,Springel:2017tpz,Nelson:2017cxy,Naiman,Marinacci:2017wew}. These simulations are different in many aspects such as the numerical methods used, the model of subgrid baryonic physics and, most importantly, the properties of simulated galaxies. 

In addition, in contrast to most of the previous analysis that only focuses on central galaxies (e.g.Refs.~\cite{Matthee:2016wir,10.1093/mnras/stz030}), our analysis includes all types of galaxies, especially including a significant fraction of satellite galaxies. Unlike the central galaxies, satellite subhalos can not be easily matched from the DMO simulations to the full baryonic physics hydrodynamical ones by identifying their dark matter particles. It can be as high as 30$\%$ low-mass satellites in the DMO simulations that can not find their counterparts in the hydrodynamical ones~\citet{Chaves-Montero:2015iga}. We, therefore, present a new approach to analyse the simulation data, which is different from the method used in ~\citet{Chaves-Montero:2015iga}.

This paper is organised as follows: in Section~\ref{Hydrosims}, we introduce the hydrodynamcial simulations and the galaxy catalogue used. In Section~\ref{Predisruption}, we discuss the physical relationship between a property of galaxy and a property of its host dark matter halo before major disruptions; in Section~\ref{postdis}, we discuss the impact of disruptions on the galaxy property-halo property relation; in Section~\ref{galaxyclustering}, we discuss how to efficiently model galaxy clustering using subhalos from simulations. in Section~\ref{conclusions}, we summarise and conclude this work.

\section{Hydrodynamical simulations}\label{Hydrosims}
The first simulation we adopt is the EAGLE simulation~\citet{Schaye:2014tpa,EAGLE2,McAlpine:2015tma}. Among the EAGLE suite, we use the largest simulation with a box size of $67.77{\rm Mpc}/h$ along one side. The mass resolution of gas  particle in this simulation is $1.81 \times 10^6 M_{\odot}$ and a DM particle is $9.70 \times 10^6 M_{\odot}$. We only use well resolved galaxies with $M_{\rm star}>8\times 10^7 M_{\odot}$. For the Illustris~\citet{Nelson:2015dga,Vogelsberger:2014dza} simulation, we use the highest resolution run with a box size of $75{\rm Mpc}/h$ along one side. A gas particle in this simulation is $1.3 \times 10^6 M_{\odot}$ and a DM particle is $6.3 \times 10^6 M_{\odot}$. We focus on galaxies with $M_{\rm star}>8\times 10^7 M_{\odot}$. In the IllustrisTNG suite~\cite{Nelson:2018uso,Pillepich:2017fcc,Springel:2017tpz,Nelson:2017cxy,Naiman,Marinacci:2017wew}, we use the simulation with a box size of $75{\rm Mpc}/h$ along one side, the same as Illustris. A gas particle in this simulation is $1.4 \times 10^6 M_{\odot}$ and a DM particle is $7.5 \times 10^6 M_{\odot}$. Again, we focus only on galaxies with $M_{\rm star}>8\times 10^7 M_{\odot}$. Further, the stellar (or gas) mass used in this work is the total mass of stars (or gases) that are bounded to a galaxy. 
\section{Pre-disruption}\label{Predisruption}
The first galaxy property we should look at is the total baryonic mass $M_{\rm star +\rm gas}$ of a galaxy, namely, the total masses of stars and cold gas in the galaxy. This is motivated by the recent discovery of the mass discrepancy relation~\cite{PhysRevLett.117.201101}, which finds that there is a tight empirical relation between the radial dependence of the enclosed baryonic-to-dynamical mass ratio and the baryonic acceleration. Such a relation also indicates that the total baryonic mass of a galaxy should be tightly correlated with the circular velocity of its host dark matter halo. In this work, we investigate this issue using hydrodynamic simulations. In order to simplify our analysis, we adopt the following strategy: before discussing a galaxy's property at the current time, we first examine the galaxy's property at the epoch of the peak value of the maximum circular velocities during its merger history $v_{\rm peak}$. This is to use a galaxy's property before major disruption, which can avoid any drastic changes in the properties of galaxies. We then discuss the impact of disruption on a galaxy's property later on.

\begin{figure*}
\includegraphics[width=\linewidth]{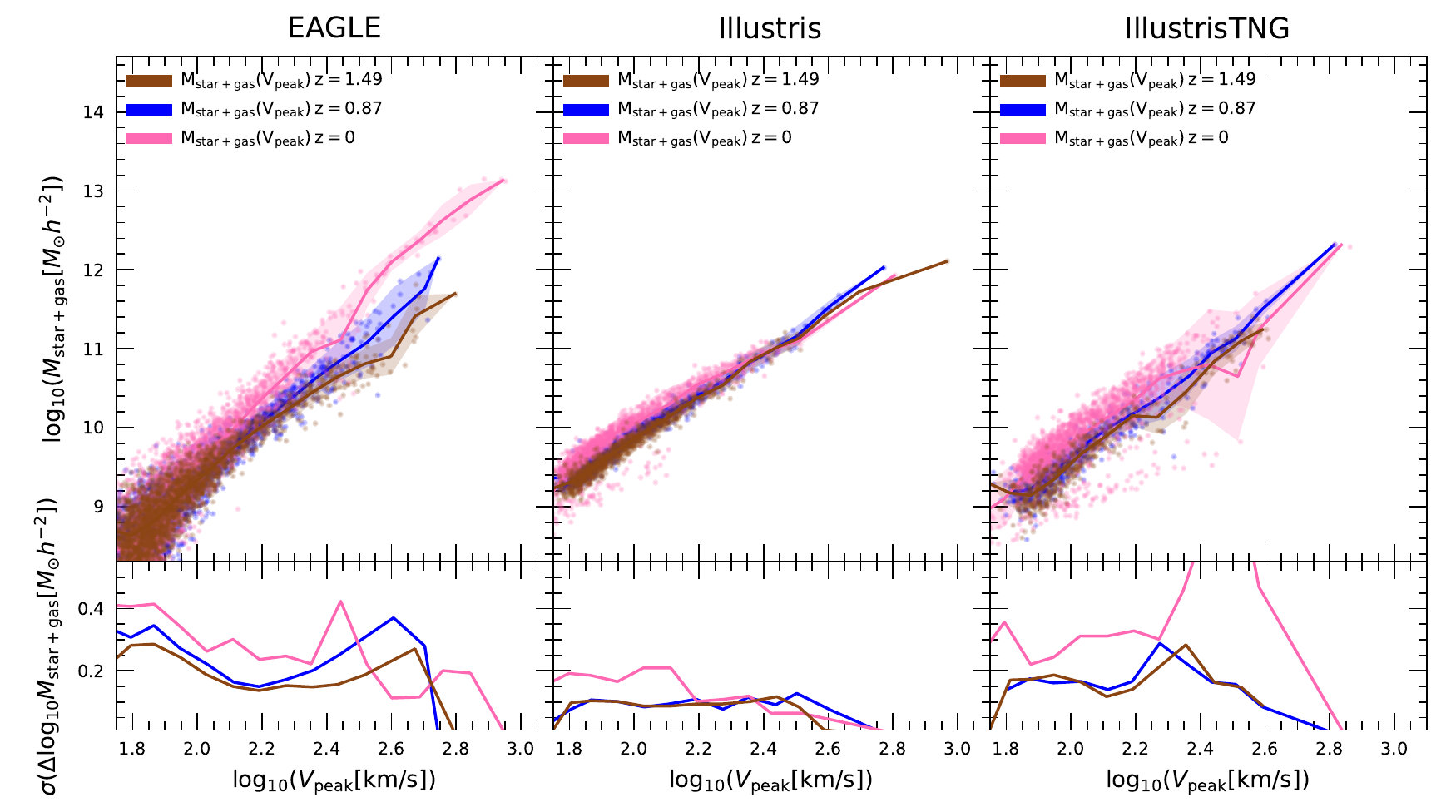}
\caption{The total baryonic mass of a galaxy (the value at the epoch of $v_{\rm peak}$) as a function of $v_{\rm peak}$ at three representative redshifts $z=1.49,0.87,0$. The samples used here include both centrals and satellites. The upper panels show the total baryonic mass-$v_{\rm peak}$ relations for three simulations EAGLE (left), Illustris (middle) and IllustrisTNG (right). The lower panels show the scatter in these relations. The EAGLE and IllustrisTNG simulations show relatively big scatter which also depends on redshift. The Illustris simulation, however, shows a smaller scatter. }\label{totalbarynoic}
\end{figure*}

The upper panels of Fig.~\ref{totalbarynoic} show the total baryonic mass of a galaxy as a function of $v_{\rm peak}$ in the EAGLE (left), Illustris (middle) and IllustrisTNG (right) simulations at three representative redshifts $z=1.49,0.87,0$. Galaxies in Fig.~\ref{totalbarynoic} are selected when their maximum circular velocities peak at the three specified redshifts during their merger histories. Hence, the value of the total baryonic mass for each galaxy in Fig.~\ref{totalbarynoic} is at $v_{\rm peak}$. Further, note that the samples used here include both centrals and satellites. From the lower panels of Fig.~\ref{totalbarynoic}, EAGLE and IllustrisTNG simulations show relatively big scatter in the total baryonic mass-$v_{\rm peak}$ relation, which is around $\sigma(\Delta \log_{10} M_{\rm star + \rm gas})\sim 0.3$. The scatter also depends on redshift. However, Illustris simulation shows a smaller scatter, which is around $\sigma(\Delta \log_{10} M_{\rm star + \rm gas})\sim 0.2$. That hydrodynamical simulations do not give a consistent result on the total baryonic mass-$v_{\rm peak}$ relation indicates that the behavior of the gas component in a galaxy is dependent on the details of baryonic physics implemented in the simulations.

\begin{figure*}
\includegraphics[width=\linewidth]{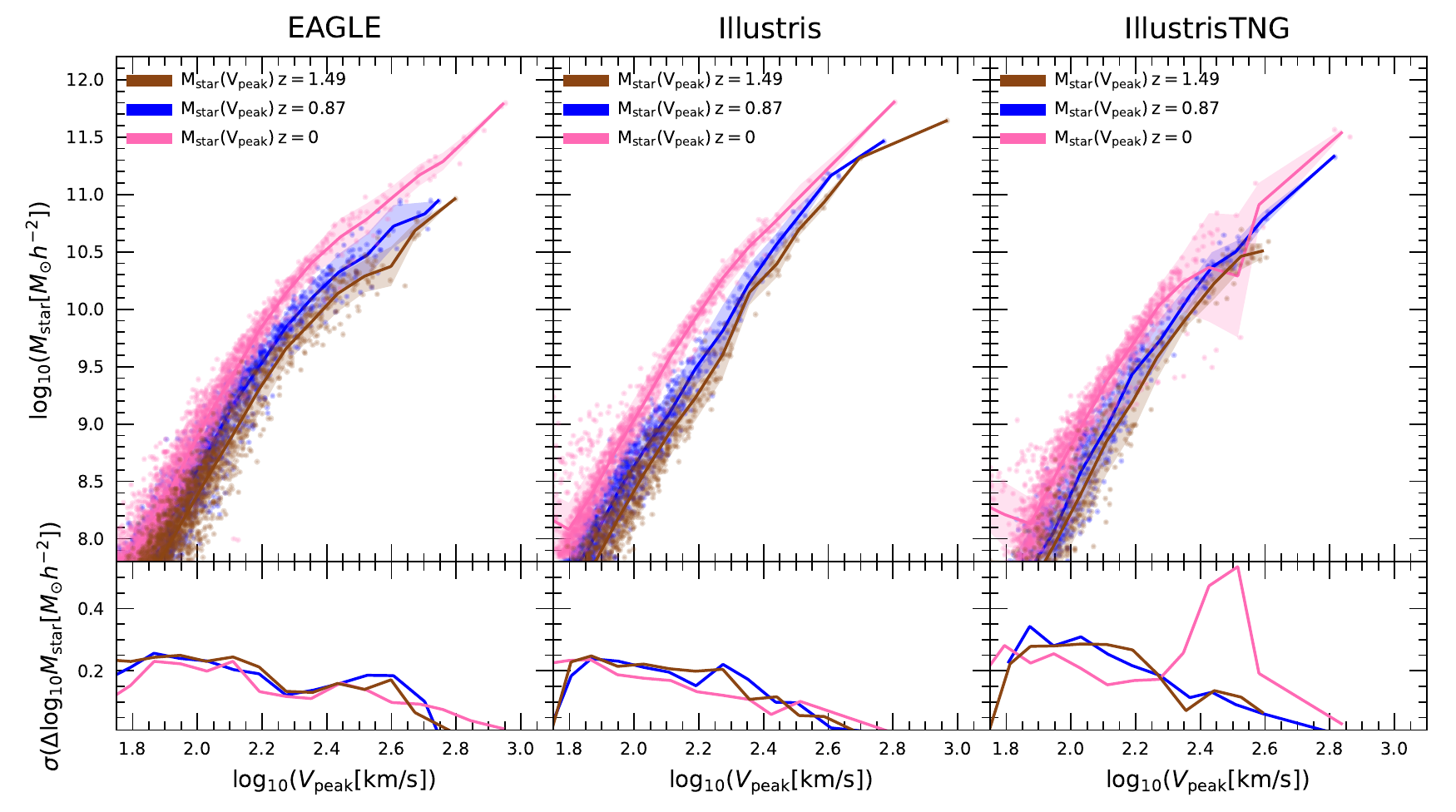}
\caption{The stellar mass of a galaxy (the value at the epoch of $v_{\rm peak}$)  as a function of $v_{\rm peak}$ at three representative redshifts $z=1.49,0.87,0$, the same as in Fig~\ref{totalbarynoic}. The samples used here include both centrals and satellites. The upper panels show  the stellar mass ($v_{\rm peak}$)-$v_{\rm peak}$ relations and the lower panels show the scatter in these relations. At a fixed redshift, the stellar mass correlates tightly with $v_{\rm peak}$ in all the three simulations and the scatter almost does not depend on redshift. The zero-point of the scaling relation varies with redshift. This is due to the evolution of stellar mass. Since a longer time of star-formation can lead to more stellar mass formed in a galaxy, the zero-points of the scaling relation is higher at low redshift than those at high redshift.}\label{stellarmass}
\end{figure*}

Next, we turn to the stellar component of a galaxy. The upper panels of Fig.~\ref{stellarmass} show the stellar mass of a galaxy (the value at the epoch of $v_{\rm peak}$) as a function of $v_{\rm peak}$ for the EAGLE (left), Illustris (middle) and IllustrisTNG (right) simulations at three different redshifts $z=1.49,0.87,0$, the same as in the previous case. In contrast to the total baryonic mass, all three simulations show a very similar tight relation between stellar mass and $v_{\rm peak}$ at a fixed redshift. From the lower panels of Fig.~\ref{stellarmass}, different simulations even at different redshifts also show a very similar size of scatter. However, the zero-points in the scaling relations are different. This is due to the fact that a galaxy's stellar mass evolves with time. A longer time of star-formation can lead to more stellar mass formed in a galaxy and the zero-points of the scaling relation is, therefore, higher at low redshift and lower at high redshift. Clearly, the evolution of the stellar mass plays an important role here. If we neglect such evolution effect and simply take the stellar mass at the epoch of $v_{\rm peak}$, the scatter in the stellar mass-$v_{\rm peak}$ relation will be very large. Therefore, the evolution of galaxy stellar mass should be accounted for. However, before we proceed to present how to correct for this effect, we first look at the specific star forming rate (sSFR), namely, the star-forming rate per unit stellar mass of a galaxy. 

\begin{figure}
\centering
\includegraphics[width=3.2in,height=3.3in]{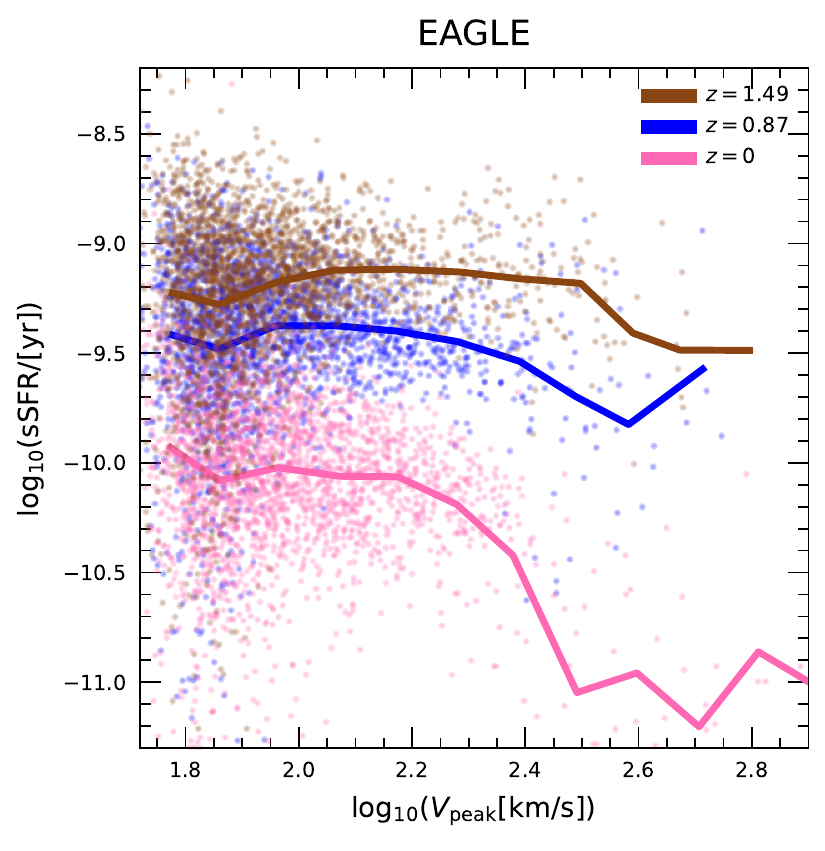}
\caption{The specific star-forming rate (sSFR) as a function of $v_{\rm peak}$ at three representative redshifts in the EAGLE simulation. Despite the big scatter around the main sequence galaxies, the average sSFR is nearly a constant, which is almost independent of $v_{\rm peak}$, except the most massive ones at redshift zero. A very similar result can be found in the Illustris and IllustrisTNG simulations as well. However, we do not present here for simplicity. }\label{sSFR}
\end{figure}

Figure~\ref{sSFR} shows the sSFR of a galaxy as a function of $v_{\rm peak}$ at three representative redshifts in the EAGLE simulation. Around the mean values (represented by solid lines and usually called the main sequence galaxies), the scatter of the sSFR is very big. However, the average value of the sSFR is nearly a constant at a fixed redshift, which is almost independent of $v_{\rm peak}$, except the most massive ones at redshift zero. Thus, one can model such evolution of stellar mass as 
\begin{equation}
\log_{10}(M_{\rm star}[M_{\odot}h^{-2}])|_{z=0}=\log_{10}(M_{\rm star}[M_{\odot}h^{-2}])(z)+\alpha\times z,\label{ecorrect}
\end{equation}
where $\alpha$ is a constant, which is different between the EAGLE, Illustris and IllustrisTNG simulations ($\alpha_{\rm EAGLE} = 0.324\, ,\,\alpha_{\rm Illustris} = 0.243, ,\,\alpha_{\rm IllustrisTNG} = 0.194$). This is because the stellar mass functions and the star forming rates in these simulations are different. The stellar mass-$v_{\rm peak}$ relation at different redshifts then can be corrected to redshift zero using the above equation.

\begin{figure*}
\centering
\includegraphics[width=\linewidth]{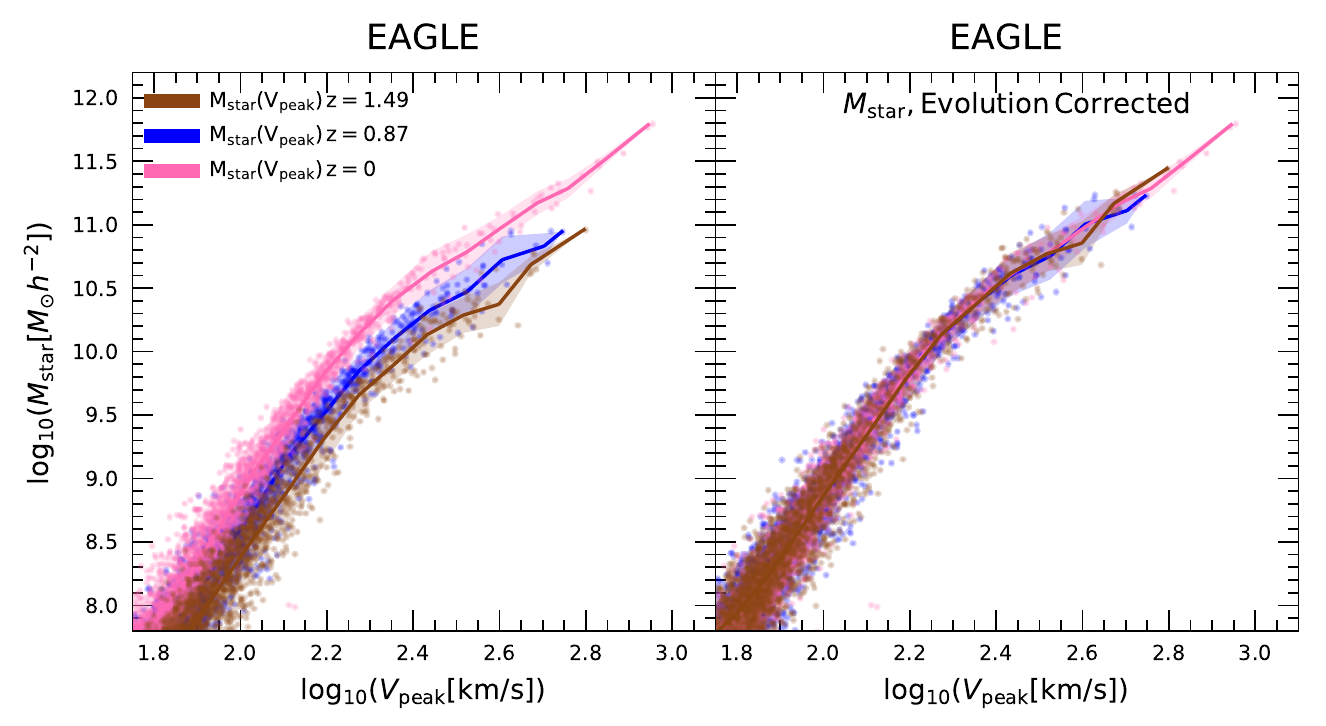}
\caption{Right: The evolution corrected stellar mass-$v_{\rm peak}$ relation at redshift zero. Left: for comparison, the stellar mass-$v_{\rm peak}$ relation at three different redshifts $z=1.49,0.87,0$, which are the same as the left panel of Fig.~\ref{stellarmass}. We only show the result from the EAGLE simulation for illustrative purposes.}\label{ecorrected}
\end{figure*}

Figure~\ref{ecorrected} shows the evolution corrected stellar mass-$v_{\rm peak}$ relation at redshift zero. We only use the EAGLE simulation here for illustrative purposes. In Fig.~\ref{ecorrected}, we also show the results at three representative redshifts as control. After taking into account the effect of stellar mass evolution, the evolution corrected stellar mass of  galaxies are tightly correlated with $v_{\rm peak}$ (right panel in Fig.~\ref{ecorrected}). The intrinsic scatter of this relation is very small. Note that here we have used the galaxy's stellar mass that would be at redshift zero if there were no disruption rather than the true current stellar mass of the galaxy. 

In addition, it is worth noting that although the scatter in the sSFR-$v_{\rm peak}$ relation is very large (see Fig.~\ref{sSFR}), the stellar mass and $v_{\rm peak}$ end up with a very tight correlation. This is indeed not a surprise because the sSFR has big fluctuations during the accretion history of stellar mass~\citet{10.1093/mnras/stz030}, which leads to the big scatter in the sSFR-$v_{\rm peak}$ relation. However, in the stellar mass-$v_{\rm peak}$ relation, this effect can be largely smoothed out as the stellar mass is the temporal cumulation of the sSFR. 

\section{Post-disruption}\label{postdis}

\begin{figure*}
\centering
\includegraphics[width=\linewidth]{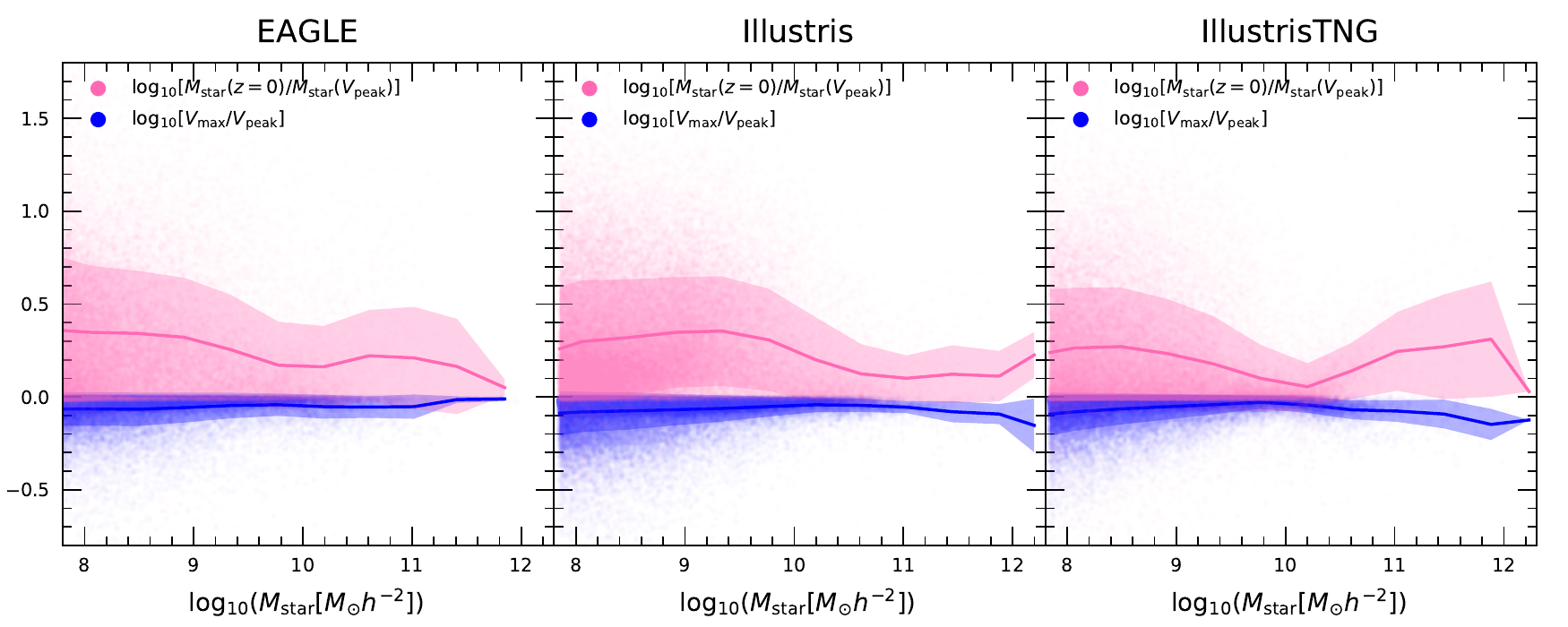}
\caption{The ratio of the stellar mass (pink points) and the maximum circular velocity $v_{\rm max}$ (blue points) of a galaxy at the present day to those values at the epoch of $v_{\rm peak}$ as a function of stellar mass for the EAGLE (left), Illustris (middle) and IllustrisTNG (right) simulations. As a significant fraction of galaxies have $\log_{10}(V_{\rm max}/V_{\rm peak})$ much less than zero (blue points), which indicates that the disruption of the dark matter component of galaxies is prevalent. In contrast, for most galaxies, the stellar component at the present day is much greater than the value at the epoch of $V_{\rm peak}\,$($\log_{10}[M_{\rm star}(z=0)/M_{\rm star}(V_{\rm peak})]>0$), which means that even the dark matter and gas components undergo disruption, star forming in most galaxies can still last for a quite long time. However, it is also evident that in some extreme cases, the stellar components suffer dramatic losses along with their dark matter components (pink points with $\log_{10}[M_{\rm star}(z=0)/M_{\rm star}(V_{\rm peak})]\ll 0$).}\label{postdisruption}
\end{figure*}

Unlike the pre-disruption stellar mass-$v_{\rm peak}$ relation, after $v_{\rm peak}$ great complexity comes in due to the interactions between galaxies. In order to illustrate this, in Fig.~\ref{postdisruption}, we show the ratio of the stellar mass (pink points) and the maximum circular velocity $v_{\rm max}$ (blue points) of a galaxy at the present day to those values at the epoch of $v_{\rm peak}$ as a function of the stellar mass for the EAGLE (left), Illustris (middle) and IllustrisTNG (right) simulations. As shown by the blue points, the values of $\log_{10}(V_{\rm max}/V_{\rm peak})$ for a significant fraction of galaxies are much less than zero, which means that the disruption of the dark matter component of the galaxies is prevalent (in most cases $v_{\rm max}$ at the present day is much smaller than the value at the epoch of $v_{\rm peak}$). In contrast, the stellar component of most galaxies (pink points), on the other hand, grows rather than decreases significantly from the epoch of $V_{\rm peak}\,$ to the present day. This happens even for the cases that the dark matter and gas components undergo a dramatic loss. This indicates that even when a galaxy becomes a satellite merging into a large system, galaxies in most cases can still continue forming stars for a long time. Only in very extreme cases, the stellar components can suffer dramatic losses along with their dark matter components due to disruptions (pink points with $\log_{10}[M_{\rm star}(z=0)/M_{\rm star}(V_{\rm peak})]\ll0$). 

\begin{figure*}
\centering
\includegraphics[width=\linewidth]{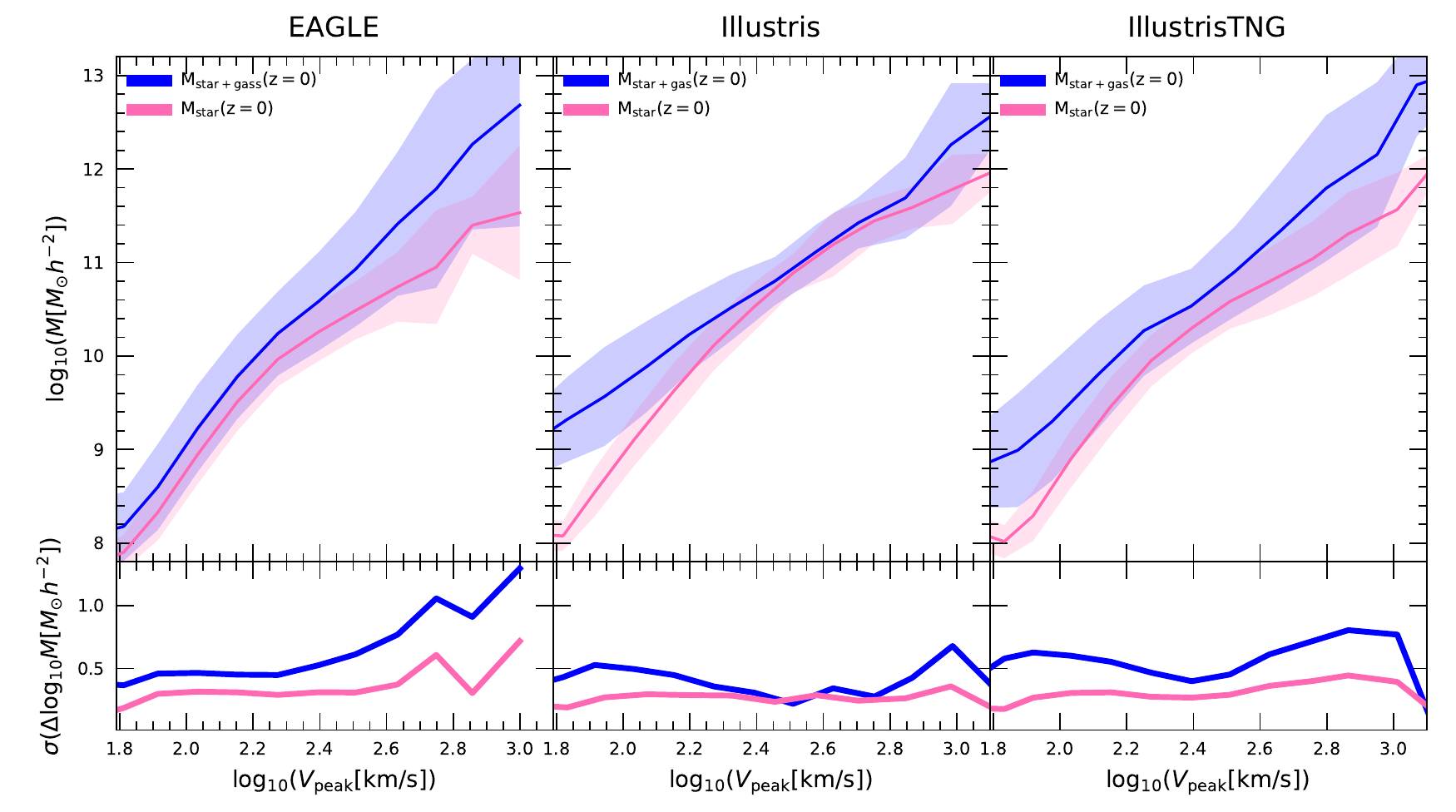}
\caption{The upper panels show the post-disruption total baryonic mass-$V_{\rm peak}$ relation, compared with the post-disruption stellar mass-$V_{\rm peak}$ relation for different hydro-dynamical simulations. 
In contrast to Fig.~\ref{totalbarynoic} and Fig.~\ref{stellarmass}, the values of the total baryonic mass and stellar mass of galaxies in this plot are taken at redshift zero. The left panels are for the EAGLE simulation, the middle ones are for the Illustris simulation and the right ones are for the IllustrisTNG simulations. The lower panels show scatters for the above relations. All hydro-dynamical simulations consistently show that in the post-disruption case, the stellar mass-$V_{\rm peak}$ relations are tighter than those in the total baryonic mass-$V_{\rm peak}$ relation.} \label{bar2star}
\end{figure*}

Figure~\ref{bar2star} shows the post-disruption total baryonic mass-$V_{\rm peak}$ relation, compared with the post-disruption stellar mass-$V_{\rm peak}$ relation for different hydro-dynamical simulations. The left panels are for the EAGLE simulation, the middle ones are for the Illustris simulation and the right ones are for the IllustrisTNG simulations. The lower panels of Figure~\ref{bar2star} show scatters for the above relations. After the epoch of $V_{\rm peak}$, both stars and gases suffer disruption. However, since there is only gravitational interaction between stars and there are significant non-gravitational interactions between gases(e.g. energetic feed-backs), gas is more prone to disruption than that of stars. The stellar mass-$V_{\rm peak}$ relation, therefore, is much tighter than that of the total baryonic mass-$V_{\rm peak}$ relation. Further, all three hydro-dynamical simulations yield a consistent result.

\begin{figure*}
\centering
\includegraphics[width=\linewidth]{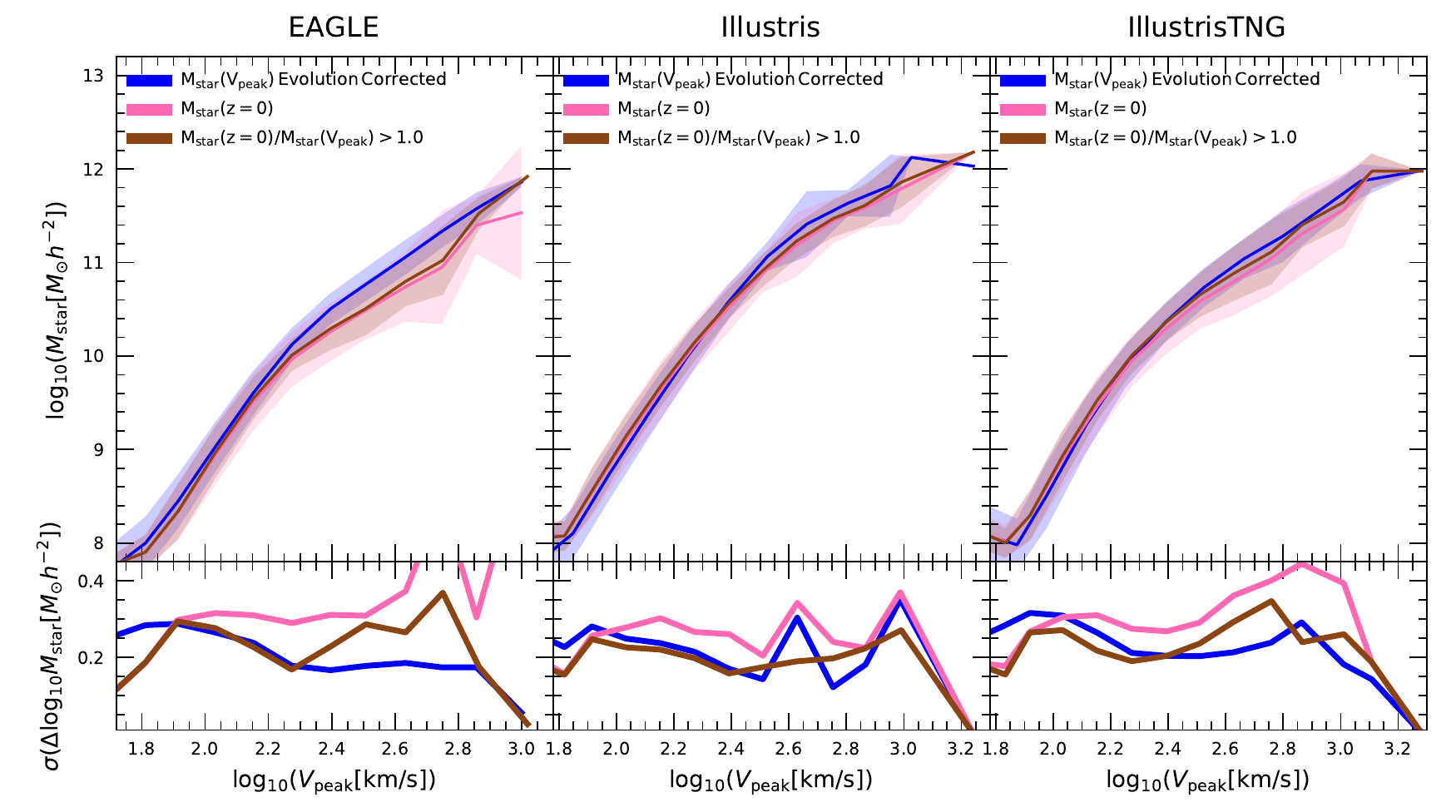}
\caption{The upper panels show the evolution corrected stellar mass-$V_{\rm peak}$ relation, in which the stellar mass is taken at the epoch of $V_{\rm peak}$ but evolution corrected to redshift zero (blue), the current stellar mass (post-disruption)-$V_{\rm peak}$ relation (pink) and the current stellar mass-$V_{\rm peak}$ relation but without disrupted galaxies $\log_{10}[M_{\rm star}(z=0)/M_{\rm star}(V_{\rm peak})]>0$ (brown). The left panels are for the EAGLE simulation, the middle ones are for the Illustris simulation and the right ones are for the IllustrisTNG simulations. The lower panels show scatters for the various relations.} \label{stellarmassvpeak}
\end{figure*}

Figure~\ref{stellarmassvpeak} shows various stellar mass-$V_{\rm peak}$ relations for the EAGLE (left), Illustris (middle) and IllustrisTNG simulations. We show the evolution corrected relation (blue), in which stellar mass is taken at the epoch of $V_{\rm peak}$ but evolution corrected to redshift zero using Eq.~(\ref{ecorrect}). In the same plot, we also present the current stellar mass (post-disruption)-$V_{\rm peak}$ relation (pink) and the current stellar mass-$V_{\rm peak}$ relation but without disrupted galaxies $\log_{10}[M_{\rm star}(z=0)/M_{\rm star}(V_{\rm peak})]>0$ (brown). The lower panels show scatters in the various relations. The typical scatter on stellar mass in the post-disruption stellar mass-$V_{\rm peak}$ relation is around $\sigma(\Delta\log_{10}M_{\rm star})\sim0.3$ in the EAGLE and IllustrisTNG simulations but slightly smaller $\sigma(\Delta\log_{10}M_{\rm star})\sim0.25$ in the Illustris simulation. As in our analysis we include both centrals and satellites, the scatters here are slightly bigger than the values reported in the previous work that only use centrals (e.g. Refs.~\cite{Matthee:2016wir,Wechsler:2018pic}). However, if we do not take into account the disrupted galaxies and only consider the galaxies that gain stellar mass after $V_{\rm peak}$, namely, $\log_{10}[M_{\rm star}(z=0)/M_{\rm star}(V_{\rm peak})]>0$, the scatter can be significantly reduced blow $\sigma(\Delta\log_{10}M_{\rm star})<0.2$ for galaxies with $V_{\rm peak}>100[\rm km/s]$ (brown). In this case, the scatter is close to the ideal case of the evolution corrected value using stellar mass at the epoch of $V_{\rm peak}$ (comparing the blue and brown lines in the lower panels of Fig.~\ref{stellarmassvpeak} ). This indicates that after $V_{\rm peak}$ when a galaxy is accreted by a larger system, stellar mass striping is a major source of scatter to the relation between the current stellar mass and $V_{\rm peak}$. Therefore, the tightest relation between a galaxy property and a halo property can only be achieved in terms of quantities before this disruption. 

\begin{figure}
\includegraphics[width=\linewidth]{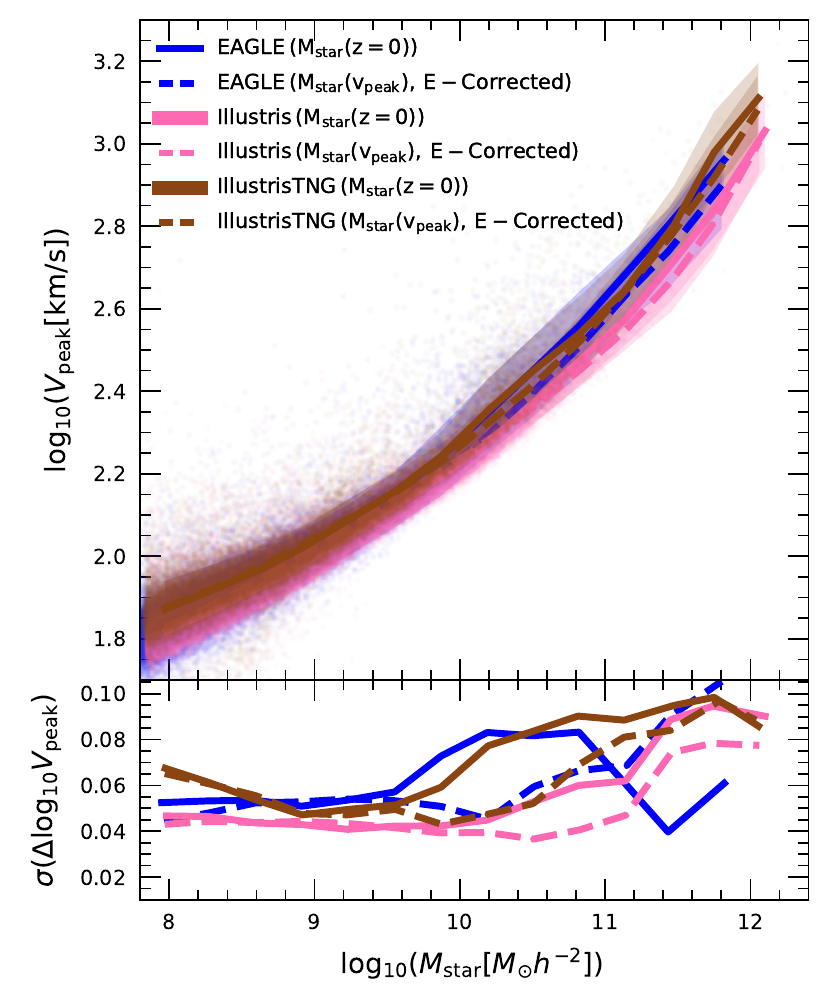}
\caption{Upper panels: comparisons of the $V_{\rm peak}$-evolution corrected stellar mass relation (solid lines) and $V_{\rm peak}$ - current stellar mass relation (dashed lines) derived from different hydrodynamical simulations. Lower panels: scatter on $\log_{10}V_{\rm peak}$.}\label{vpeakmass}
\end{figure}

Nevertheless, it is worth pointing out that even in the case of post-disruption, the current stellar mass and $V_{\rm peak}$ still exhibit a very tight correlation. In Fig.~\ref{vpeakmass}, we compare such a relation (solid lines) with the evolution corrected value (dashed lines). Unlike Fig.~\ref{stellarmassvpeak}, in Fig.~\ref{vpeakmass} we show the scatter on $\log_{10}V_{\rm peak}$. This has more practical meaning as in observations we can only measure a galaxy's current stellar mass (post-disruption). Given the observed galaxy catalogue, we usually add scatter in the $V_{\rm peak}$-ranked subhalo catalogues in $N$-body simulations and then compare them with observations. From Fig.~\ref{vpeakmass}, even considering the post-disruption galaxies, the scatter on $\log_{10}V_{\rm peak}$ is still very small $\sigma(\Delta\log_{10}V_{\rm peak})<0.08$, only except the most massive cases, which, however, account for only a small fraction of the total galaxies. 

\begin{figure*}
\centering
\includegraphics[width=3.4in,height=3.4in]{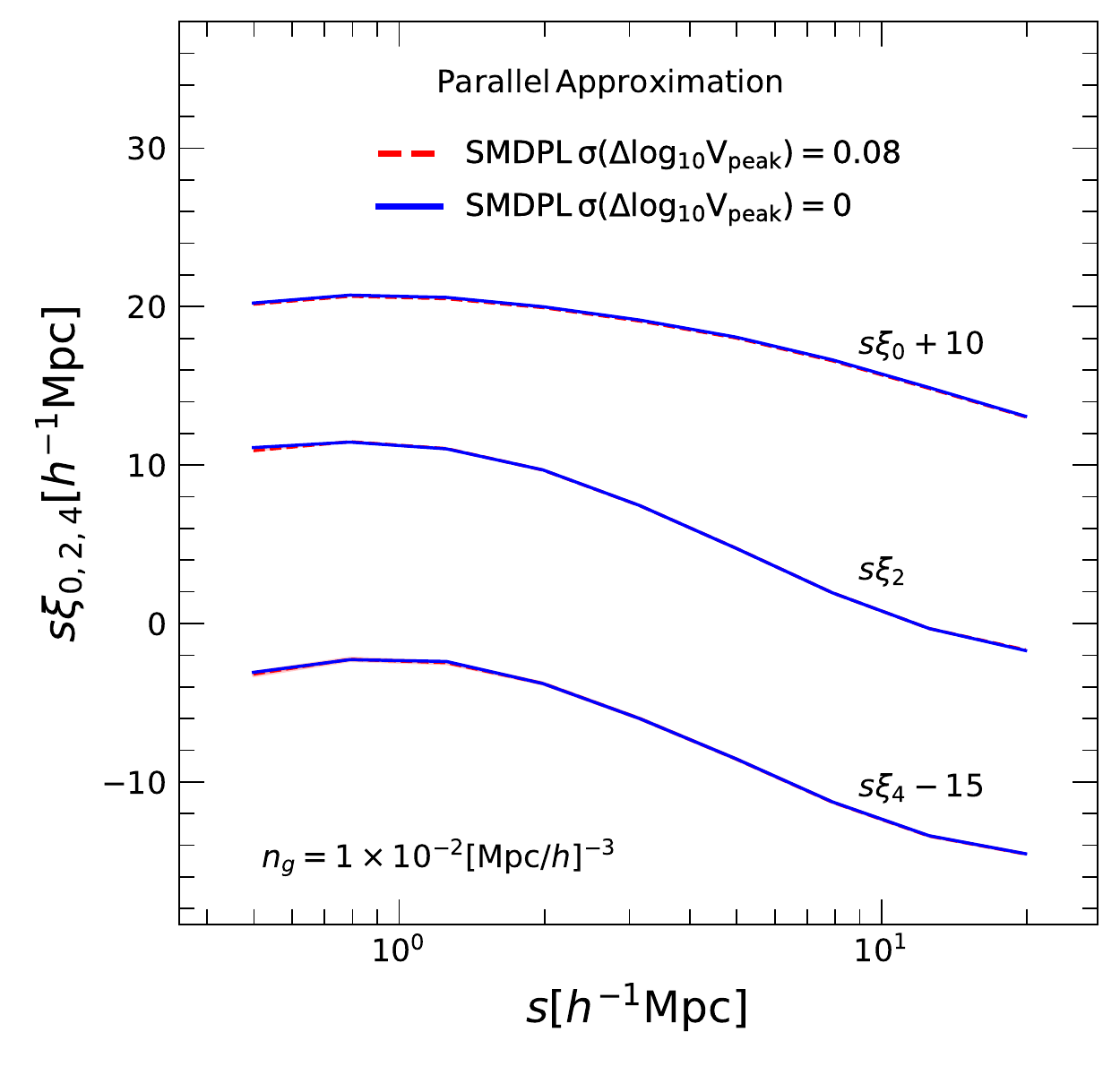}
\includegraphics[width=3.4in,height=3.4in]{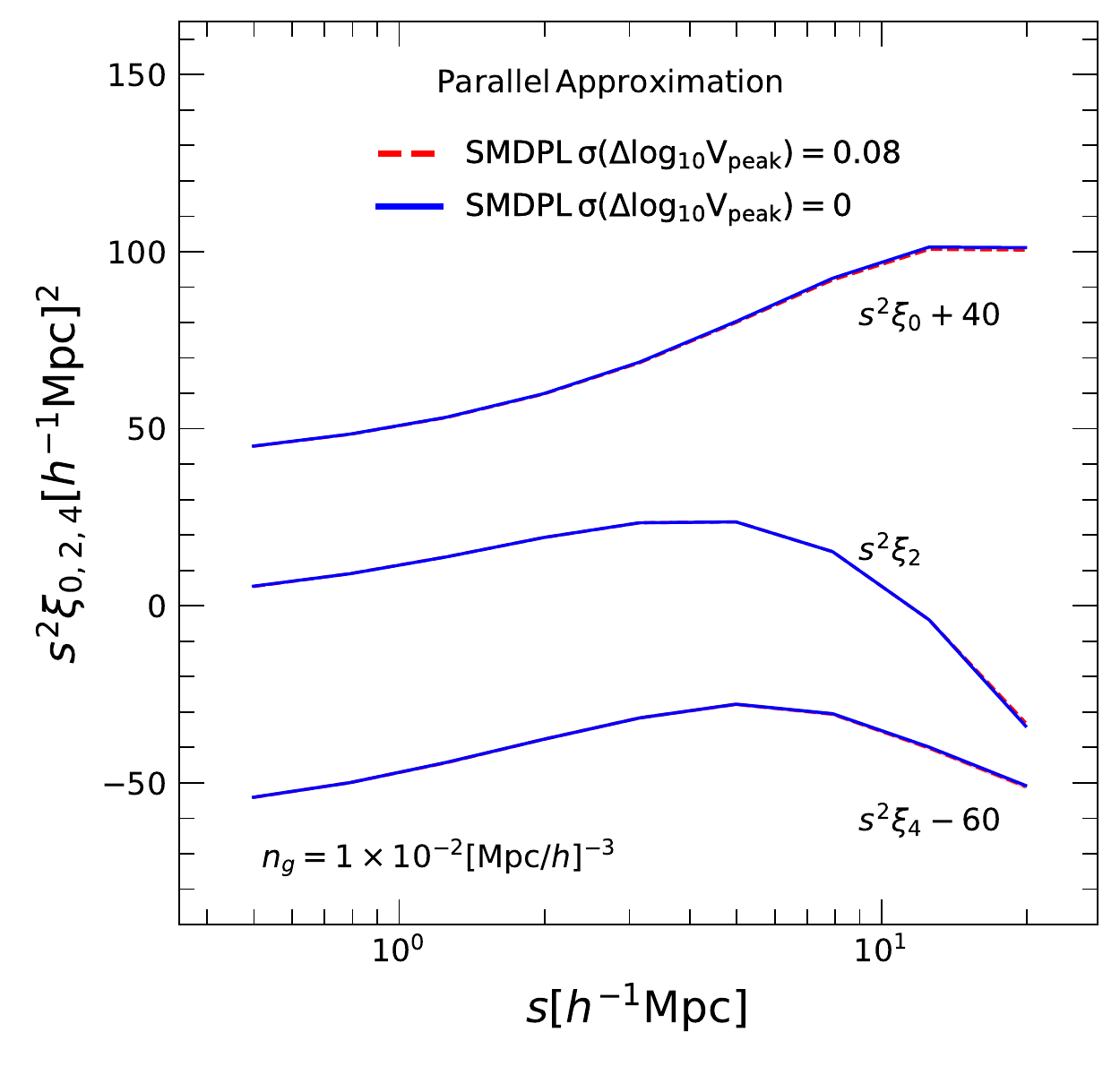}
\caption{Left: The impact of scatter on the predicted RSD multipoles. The solid lines are for the results without scatter. The dashed lines are for the results with scatter. The shaded regions on the dashed lines are the 1$\sigma$ uncertainty around the expectation value, derived from 200 realisations. The shaded regions are so small that they can not be clearly seen in the plot. The RSD multipoles $\xi_{0,2,4}$ are multiplied by the redshift space separation $s$. This rearrangement makes the RSD multipoles more sensitive to changes at small scales. Right: Similar to the left, but the RSD multipoles $\xi_{0,2,4}$ are multiplied by $s^2$. The larger index of $s$ can make the RSDs multipoles more sensitive to changes at large scales. }\label{Fig:assembely}
\end{figure*}

\begin{figure}
\centering
\includegraphics[width=3.3in,height=3.3in]{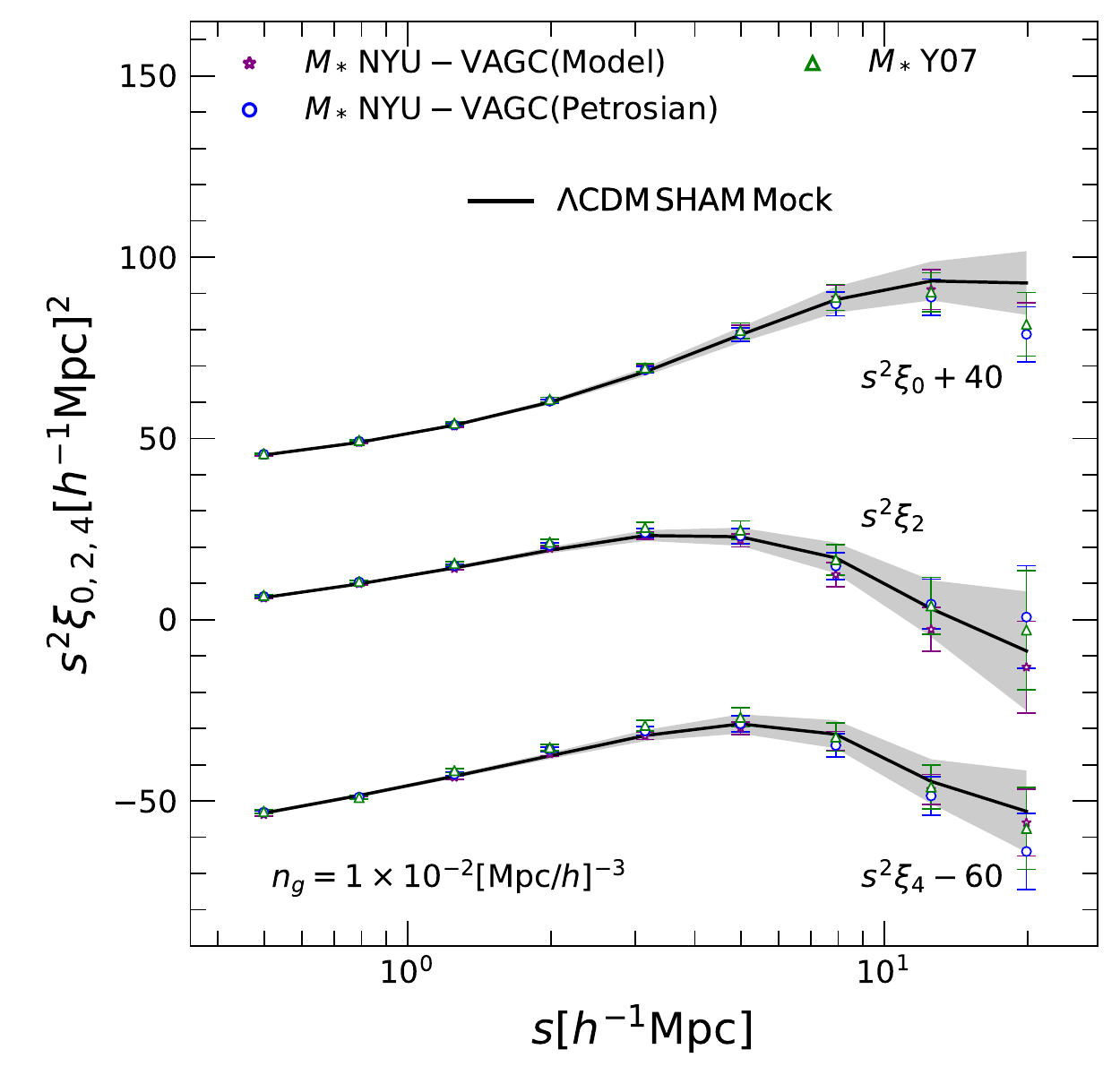}
\caption{Redshift-space multipoles (monopole $\xi_0$, quadrupole $\xi_2$, and hexadecapole $\xi_4$) for subhalos (solid black lines) and the measurements from the SDSS main galaxy samples (symbols with error bars). The galaxy samples are constructed as volume-limited and complete in stellar mass. The number density is chosen as $n=0.01[{\rm Mpc}/h]^{-3}$. The predictions from subhalos agree impressively well with the observations. Note that the subhalos are simply selected by ranking $V_{\rm peak}$. So there is no free parameter in our mock galaxy catalogue. In addition, we also show the observational measurements obtained using three different stellar mass estimators: a template-fit method as adopted in the NYU catalogue with the SDSS model magnitudes (stars), the same template-fit method but with SDSS Petrosian magnitudes (circles), and a single-color method (triangles). The black shaded regions represent the 1$\sigma$ uncertainty in the theoretical predictions and the error bars are derived from 133 realisations using the jack-knife re-sampling technique.}\label{Fig:SDSS}
\end{figure}

\section{modelling galaxy clustering}\label{galaxyclustering}
The existence of such a tight correlation between galaxy stellar mass and $V_{\rm peak}$ has an important implication on how to efficiently model galaxy clustering from N-body simulations. As demonstrated in the previous sections, even for hydrodynamical simulations with very different subgrid baryonic physics and numerical methods used, they still produce similar predictions on such relation (see Fig.~\ref{vpeakmass}). Therefore, the predicted galaxy stellar mass-$V_{\rm peak}$ relation should be robust unless there are serious limitations in our current understanding of galaxy formation. Given such a tight correlation, galaxy clustering should be able to be modeled by using $V_{\rm peak}$-selected subhalos, which corresponds to stellar mass selected galaxies in the real observations. In this section, we will test this point with observations. 

Before comparing simulations with observations, we first test the impact of scatter in the stellar mass-$V_{\rm peak}$ relation on the clustering of subhalos. In our analyses, we use the Small MultiDark Planck simulation (SMDPL)~\cite{Klypin:2014kpa}, which uses $3840^3$ dark matter particles in a box of $400{\rm Mpc}/h$ along one side.  We focus on the redshift space since the redshift space clustering can be directly measured in the real observations. We add scatter in the halo catalogue as follows. We first take subhalos from the original catalogue and then replace their values of $V_{\rm peak}$ by drawing a random number around the logarithm of $V_{\rm peak}$ ( $\log_{10}V_{\rm peak}$) within 1$\sigma(\Delta\log_{10}V_{\rm peak})=0.08$ scatter. Then we re-rank all the subhalos in the catalogue and generate 200 realisations for such catalogues. The choice of the value of scatter is motivated by Fig.~\ref{vpeakmass}, from which $\sigma(\Delta\log_{10}V_{\rm peak})=0.08$ is large enough for most galaxies only except the most massive ones. The most massive galaxies, however, only account for a small fraction of the total number of galaxies.

In order to measure the redshift-space two point correlation function $\xi(r_{\sigma},r_{\pi})$, we use the Landy and Szalay estimator~\cite{LSestimator} 
\begin{equation}
\xi(r_{\sigma},r_{\pi})=\frac{DD(r_{\sigma},r_{\pi})-2DR(r_{\sigma},r_{\pi})+RR(r_{\sigma},r_{\pi})}{RR(r_{\sigma},r_{\pi})}\quad,\label{L_S}
\end{equation}
where $DD$, $DR$ and $RR$ are the data-data, data-random, random-random pair counts and $r_{\sigma}$, $r_{\pi}$ are the separations of galaxy pairs perpendicular and parallel to the line-of-sight direction, respectively.  $\xi(r_{\sigma},r_{\pi})$ can be expanded in terms of Legendre polynomials
\begin{equation}
\xi_l(s)=\frac{2l+1}{2}\int_{-1}^{1}d\mu \xi(s,\mu)P_l(\mu)\, ,\label{Intxi}
\end{equation}
where $P_l(\mu)$ is the Legendre polynomial of order $l$, $s=\sqrt{r_{\sigma}^2+r_{\pi}^2}$ and $\mu=r_{\pi}/s$. We use logarithmic bins in $s$ ($d\log_{10}s = 0.2$) and measure $s$ up to $s_{\rm max} = 25 {\rm Mpc}/h$. We numerically work out the integration in Eq.~(\ref{Intxi}) with $d\mu = 0.05$.

Figure~\ref{Fig:assembely} shows the impact of scatter on the predicted multipoles of redshift space distortions (RSD) of subhalos with a number density of $n=0.01[{\rm Mpc}/h]^{-3}$. The solid lines are for the results without scatter. The dashed lines are for the results with scatter. The left panel of Fig.~\ref{Fig:assembely} shows the RSD multipoles $\xi_{0,2,4}$ multiplied by the redshift space separation $s$, which makes the RSD multipoles more sensitive to changes at small scales. The right panel shows similar results but for RSD multipoles $\xi_{0,2,4}$ multiplied by $s^2$. The larger index of $s$ in this case can make the RSD multipoles more sensitive to changes at large scales. Overall, in both cases, the scatter has a very limited impact on the RSD multipoles of subhalos. This result indeed can be expected. Given the high number density of our samples, the subhalos that would be in and out of the catalogue due to scatter only account for a small fraction of the total samples. The clustering of subhalos, therefore, is stable against such scatter. Further, note that in the above analysis, we have adopted the parallel approximation to add the RSD effects for subhalos in simulations.

Figure~\ref{Fig:SDSS} shows the predicted multipoles (monopole $\xi_0$, quadrupole $\xi_2$, and hexadecapole $\xi_4$) of RSDs (black solid lines) compared to the measurements from the SDSS data (symbols with error bars), which is volume-limited and complete in galaxy stellar mass with a number density of $n=0.01[{\rm Mpc}/h]^{-3}$ (see Ref.~\cite{He:2018oai} for details). The predictions from subhalos agree impressively well with observations even at relatively large scales $r\sim 20 {\rm Mpc}/h$. Note that here we use a realistic mock catalogue by collating 8 replicas of the box and place the observer at the centre. This mock catalogue has the same survey mask as the real data. The subhalos are simply selected by ranking $V_{\rm peak}$ from the SMDPL simulation. So there is no free parameter in our mock catalogue. In order to demonstrate the robustness of our RSD measurements, we test three different estimators of stellar masses: a template-fit method originally adopted in the NYU catalogue with the SDSS model magnitudes (stars)~\cite{Blanton:2004aa}, the same template-fit method but using SDSS Petrosian magnitudes (circles), and a single-color method (triangles)~\cite{Yang:2007yr}. 

\section{Conclusions}\label{conclusions}
\label{sec:conclusions}
In this work, we have investigated the galaxy property-halo property relation using hydrodynamical simulations of galaxy formation. Unlike the conventional phenomenological frameworks such as HOD or CLF, the advantage of hydrodynamical simulations is that it can provide a clear physical picture about how a property of a galaxy relates to a property of its host dark matter halo. Based on state-of-the-art hydrodynamical simulations, such as EAGLE, Illustris and IllustrisTNG, we study under what circumstance a galaxy's property could correlate most tightly with a property of its host dark matter halo. In addition, in contrast to most of the previous work,  our analyses include all types of galaxies, especially containing a significant fraction of satellites. Unlike the central galaxies, satellite subhalos can not be easily matched from the DMO simulations to the full baryonic physics hydrodynamical ones. 
Therefore, we adopt a novel approach that is different from the one used in Ref.~\citet{Chaves-Montero:2015iga} to analyse the simulation data. The main findings of our work are summarised as follows:

\begin{itemize}
\item Despite the presence of the observed mass discrepancy relation, hydrodynamical simulations do not show a consensus on the tight correlation between the total baryonic mass of a galaxy and the circular velocity of its host dark matter halo if considering all types of galaxies at the present time (post-disruption) (see Fig.~\ref{bar2star}). Within $\Lambda$CDM, this can be expected as the gas component of a galaxy is strongly affected by various feed-backs in the processes of galaxy formation in the first place. The gas component can also have non-gravitational interactions and, therefore, is more prone to disruptions than that of the dark matter and stellar components since the latter two only have gravitational interactions among them. Therefore, such a tight correlation can not be expected to exist in satellite galaxies . 

\item The stellar mass of a galaxy at the epoch of $V_{\rm peak}$ with an evolution correction correlates most tightly to $V_{\rm peak}$ of the galaxy (see Fig.~\ref{ecorrected}).

\item After accretion, star formation in most galaxies can last for a quite long time. Most galaxies, therefore, can still gain significant amounts of stellar mass (see Fig.~\ref{postdisruption}). The stellar mass evolution reduces rather than increases the scatter in the current stellar mass-$V_{\rm peak}$ relation.

\item Aside from the intrinsic scatter, star stripping is the main cause for the scatter in the current stellar mass (post-disruption)-$V_{\rm peak}$ relation (see Fig.~\ref{stellarmassvpeak}).

\item Even for the current stellar mass (post-disruption)-$V_{\rm peak}$ relation, hydrodynamical simulations still predict a very small scatter on $\log_{10}V_{\rm peak}$ (see Fig.~\ref{vpeakmass}), which means that the clustering of stellar mass selected galaxies in observations can be well modeled by $V_{\rm peak}$-selected subhalos in simulations.

\item Since the scatter predicted by hydrodynamical simulations has a very limited impact on the clustering of dense $V_{\rm peak}$-selected subhalos (see Fig.~\ref{Fig:assembely}), even simple subhalo abundance matching (without scatter and free parameter) can yield a robust prediction of galaxy clustering. We show that when compared with the SDSS main galaxy samples, the predictions of SHAM mock catalogue agree impressively well with the observations (see Fig.~\ref{Fig:SDSS}).  

\end{itemize}

It is worth noting that, in the above conclusions, we have assumed that baryons have a limited impact on the positions and motions of subhalos. This assumption is based on the findings reported in Refs.~\cite{Chaves-Montero:2015iga,Hellwing:2016ucy,McAlpine:2015tma}. The authors there have compared the clustering of subhalos that can be well matched from DMO simulations to the hydrodynamical ones with the clustering of the simulated galaxies. They find that the effect of baryons is very small on scales greater than $>1{\rm Mpc}/h$. However, an important issue here is that Refs.~\cite{Chaves-Montero:2015iga,Hellwing:2016ucy} only focus on the matched subhalos. But in fact a significant fraction of subhalos in DMO runs can not find their counterparts in the hydrodynamical ones in the first place as dark matter substructures are prone to disruptions due to numerical errors. It is unclear yet how to gauge the effect of baryons for those un-matched subhalos. Moreover, as recently pointed out by Refs.~\cite{vandenBosch:2018tyt,vandenBosch2017}, even modern state-of-the-art DMO simulations might not reliably resolve the dark matter substructures due to the insufficient force resolution in simulations. So to exactly pin down the physical effect  of baryons, rather than numerical artifacts, on the clustering and motions of dark matter substructures are indeed highly non-trivial. A detailed analysis is needed, which, however, is beyond the scope of this work and will be explored in our future work.

\section*{Acknowledgements}
We acknowledge the Virgo Consortium for making their simulation data available. The eagle simula- tions were performed using the DiRAC-2 facility at Durham, managed by the ICC, and the PRACE facil- ity Curie based in France at TGCC, CEA, Bruy`eres- le-Chˆatel. J.H.H. acknowledges support of Nanjing University and this work used the DiRAC@Durham facility managed by the Institute for Computational Cosmology on behalf of the STFC DiRAC HPC Facility (www.dirac.ac.uk). The equipment was funded by BEIS capital funding via STFC capital grants ST/K00042X/1, ST/P002293/1, ST/R002371/1 and ST/S002502/1, Durham University and STFC operations grant ST/R000832/1. DiRAC is part of the National e-Infrastructure.





\bibliographystyle{mnras}
\bibliography{myref.bib} 





\bsp	
\label{lastpage}
\end{document}